\documentclass[
 reprint,
superscriptaddress,
 amsmath,amssymb,
 aps,
]{revtex4-1}

\usepackage{graphicx}
\usepackage{dcolumn}
\usepackage{bm}
\usepackage{psfrag}
\usepackage{ulem}

\begin{document}

\preprint{APS/123-QED}

\title{Traffic congestion in interconnected complex networks}

\author{Fei Tan}
\affiliation{Department of Information Science and Electronic Engineering, Zhejiang University, Hangzhou 310027, People's Republic of China}

\affiliation{Department of Electronic and Information Engineering, The Hong Kong Polytechnic University, Hung Hom, Kowloon, Hong Kong}
\author{Jiajing Wu}
\affiliation{Department of Electronic and Information Engineering, The Hong Kong Polytechnic University, Hung Hom, Kowloon, Hong Kong}
\author{Yongxiang Xia}
\email{xiayx@zju.edu.cn}
\affiliation{Department of Information Science and Electronic Engineering, Zhejiang University, Hangzhou 310027, People's Republic of China}
\author{Chi K. Tse}
\affiliation{Department of Electronic and Information Engineering, The Hong Kong Polytechnic University, Hung Hom, Kowloon, Hong Kong}

\date{\today}

\begin{abstract}
Traffic congestion in isolated complex networks has been investigated extensively over the last decade. Coupled network models have recently been developed to facilitate further understanding of real complex systems. Analysis of traffic congestion in coupled complex networks, however, is still relatively unexplored. In this paper, we try to explore the effect of interconnections on traffic congestion in interconnected BA scale-free networks. We find that assortative coupling can alleviate traffic congestion more readily than disassortative and random coupling when the node processing capacity is allocated based on node usage probability. Furthermore, the optimal coupling probability can be found for assortative coupling. However, three types of coupling preferences achieve similar traffic performance if all nodes share the same processing capacity.  We analyze interconnected Internet AS-level graphs of South Korea and Japan and obtain similar results. Some practical suggestions are presented to optimize such real-world interconnected networks accordingly.

\begin{description}
\item[DOI]  \item[PACS number(s)]
89.75.Hc, 89.75.Fb, 89.20.Hh, 89.40.-a
\end{description}
\end{abstract}

\pacs{Valid PACS appear here}
\maketitle


\section{\label{sec:level1}Introduction}

Modern society depends greatly on the efficient operation of many critical networked infrastructures such as power grids, the Internet, transportation networks and so on \cite{cui2010complex}. Traffic on such networked systems is a significant issue. A wealth of studies on traffic congestion from the perspective of an isolated complex network framework have been conducted over the past decade \cite{zhao2005onset,arenas2001communication,de2009congestion,wuanalysis}. It has been widely revealed that traffic congestion is closely related to the network structure \cite{boccaletti2006complex,zhao2005onset,guimera2002dynamical,guimera2002optimal}. Typically, Zhao et al. \cite{zhao2005onset} examined several representative topologies of complex networks and presented the corresponding theoretical estimations of the traffic capacity. Motivated by Zhao's work, two general types of scenarios have been proposed to alleviate traffic congestion and improve traffic performance, namely, modification of the network topology and design of more effective routing algorithms (see the review article \cite{chen2012traffic} and references therein). Compared with the potential cost of changing the structure of well-established networked systems, proposals of clever routing criteria seem to be more practical and thus have attracted much interest \cite{yan2006efficient,danila2006optimal,ling2010global,tan2013hybrid}. Among numerous different kinds of proposed routing protocols, the efficient routing criterion is widely acknowledged for its simplicity and efficiency \cite{yan2006efficient}. This is actually contrary to the widely used shortest path algorithm in terms of the usage of hub nodes. Additionally, given a network framework and specific routing protocol, optimization of traffic resource allocation has been shown to be a reasonable approach to mitigating traffic congestion as well \cite{xia2010optimal,xiang2013traffic,zhang2011enhancing}. Notice that in reference \cite{xiang2013traffic} authors proposed an optimal resource allocation strategy and showed analytically how the shortest path strategy can alleviate traffic congestion to the largest extent.

However, the underlying network structures in most previous investigations about traffic congestion were principally modeled and analyzed as isolated networks. On the contrary, modern infrastructures are actually coupled together and thus significantly interact with and/or depend on each other (see papers \cite{buldyrev2010catastrophic,gao2011networks,brummitt2012suppressing} and references therein). Therefore, analysis of traffic congestion in coupled complex networks is expected to enable us better to model the traffic dynamics of real-world networks and then optimize traffic performance.

Recently, coupled network models have been developed \cite{buldyrev2010catastrophic,brummitt2012suppressing,tan2013cascading,kurant2006layered,gu2011onset,gomez2013diffusion,morris2012transport}. Some features of these dynamical processes are remarkably different from those of isolated networks. Cascading failures, for example, have been of great interest to researchers over the last few years. For cascades scenario of interdependent networks, failures of nodes in one network result in collapses of counterpart ones in the other network. Such dependency between networks makes interdependent networks even vulnerable to random failures, which is absent from isolated networks \cite{buldyrev2010catastrophic}. With respect to cascades of load in interconnected networks where two competing forces of redundant capacity and propagation of failures affect network robustness, the optimal coupling preference and/or coupling probability could be found \cite{brummitt2012suppressing,tan2013cascading}. Actually, multilayered networks have also been introduced to facilitate the estimation of traffic load in real-life systems \cite{kurant2006layered}. Besides, cooperation between layered networks and more general diffusion-like processes have been studied \cite{gu2011onset,gomez2013diffusion}. Morris et al. also have revealed key features of transport processes on coupled spatial networks \cite{morris2012transport}. Nonetheless, analysis of data-packet traffic in interconnected communication networks yet remains missing. In essence, the transport efficiency of data packets in communication networks can never be overemphasized in the cyber age.

Inspired by an abundance of research on traffic congestion in isolated complex networks and the newly developed concept of interconnected networks, this paper focuses on the effect of interconnections on traffic congestion in interconnected networks based on the data-packet transport model. Given transport scenarios, we try to explore how the interplay of coupling preference and coupling probability controls traffic in interconnected Barab\'{a}si-Albert (BA) scale-free networks. Furthermore, we collect and analyze real interconnected networks composed of the Internet AS-level topologies of South Korea and Japan and then give some workable suggestions about optimization of interconnected links between two countries.


\section{\label{sec:level2}Model}

Without loss of generality, we consider the case of only two BA scale-free networks labeled as $A$ and $B$, which can capture the heterogeneity of many real-world networked systems \cite{barabasi1999emergence}. For simplicity and clarity of the results, our model is based on the assumption that these two networks are of the same size (i.e., the number of nodes $N=N_A=N_B$ ) and same average degree $\langle k \rangle=\langle k_A \rangle=\langle k_B \rangle$. These two isolated networks are connected by adding some links, which can provide paths for traffic between them. The {\it coupling probability} $P$ is defined as the ratio between the number of interconnected links $N_{il}$ and network size $N$, namely
\begin{equation}
P=\frac{N_{il}}{N}.
\end{equation}
It is also assumed that each node has one interconnected link at most. $P$ is thus in the range from $0$ to $1$. Apart from the density of interconnected links, the way in which these links are connected also have significant effects on the dynamical processes of the two coupled networks \cite{parshani2010inter,shai2012effect,tan2013cascading}. Since our interest is to observe traffic congestion caused by uneven distribution of traffic load, three different kinds of {\it coupling preferences} based on the heterogeneity of load in individual scale-free networks are described as follows \cite{tan2013cascading}.

\begin{itemize}
  \item  Assortative Coupling. Nodes are first sorted in networks $A$ and $B$ respectively, both in the descending order of load. If different nodes share the same load, we sort them at random. Connect the first node in network $A$ with the first node in network $B$, and then connect the second node in network $A$ with the second node in network $B$, and so on. Repeat this process until $N\times P$ interconnected links are added.

  \item Disassortative Coupling. Nodes are first sorted in network $A$ ($B$) in the descending (ascending) order of load.  If different nodes share the same load, we sort them at random. Connect the first node in network $A$ (with the heaviest load) with the first node in network $B$ (with the lightest load), and then connect the second node in network $A$ with the second node in network $B$, and so on. Repeat this process until $N \times P$ interconnected links are added.

  \item Random Coupling. Randomly choose a node in network $A$ and a node in network $B$. If neither of them has an interconnected link, then connect them. Repeat this process until $N \times P$ interconnected links are added.
\end{itemize}

Data packets are usually transported based on a specific routing criterion, the corresponding algorithmic betweenness can thus approximate the traffic load. To be concrete, the algorithmic betweenness of node $k$ is defined as \cite{duch2006scaling}
\begin{equation}
\label{betweenness}
B_k=\sum_{s\neq t}\frac{n_{st}^k}{g_{st}},
\end{equation}
where $g_{st}$ is the total number of possible paths from node $s$ to node $t$ according to a specific routing algorithm (including but not limited to the shortest path protocol) and $n_{st}^k$ is the number of such paths running through node $k$.

As supposed in previous concerned literature \cite{yan2006efficient, tan2013hybrid}, all nodes are treated as both routers and hosts in this paper. That is to say, every node can generate and process data packets. Given a network, packets are generated in source nodes and then delivered to their destinations. In this process, the traffic resource allocation, such as the node processing capacity and link bandwidth, has a great impact on traffic congestion. As in many previous studies \cite{ling2010global, zhang2011enhancing}, each link is assumed to have sufficient bandwidth. Therefore, the allocation of node processing capacity is our interest here. It is widely acknowledged that most physical and technical parameters of well-established critical communication devices are not allowed to change easily. Thus, we presume that the processing capacity of nodes has been determined before they are connected with each other and remains unchanged after forming the corresponding interconnected networks.

Our model adopts two typical allocation scenarios. In either case, the total processing capacity of each isolated network is supposed to be the network size $N_A$ and $N_B$, respectively. In the first scenario, the processing capacity of each node in individual networks is allocated uniformly and thus is one. This allocation strategy is widely seen in literature about traffic on complex networks \cite{yan2006efficient,ling2010global}. We thus call it UNI for short. In the other scenario, the processing capacity of each node ($C_i$ for network $A$ and $C_j$ for network $B$) is proportional to the node's algorithmic betweenness in individual networks. They can be denoted as $C_i=\frac{B_i}{\sum^{N_A}_{s=1}B_s}{N_A}$ and $C_j=\frac{B_j}{\sum^{N_B}_{s=1}B_s}{N_B}$, respectively. A similar concept named {\it node usage probability} can be used here to represent the ratio between the algorithmic betweenness of a node and the total algorithmic betweenness in each network \cite{wuanalysis}. Therefore, $C_i=U_A(i){N_A}$ and  $C_j=U_B(j){N_B}$, where $U_A(i)$ and $U_B(j)$ are the node usage probability of node $i$ in network $A$ and of node $j$ in network $B$, respectively. From the perspective of mitigating traffic congestion, it has been proved to be the best resource allocation strategy \cite{xiang2013traffic}. We can call it NUP for short. In the simulation, for node $k$ with $C_k < 1$, it is assumed to process a packet with probability $C_k$ per time step. In addition, if the processing capacity of node $k$ is, for example, $C_k = 1.4$, it processes two packets with probability 0.4 and otherwise one packet per time step \cite{zhang2011enhancing}.

Due to the finite processing capacity, a queue of buffers is needed at each node to accommodate packets waiting for being processed. We assume that each buffer has a sufficient length, and first-in-first-out (FIFO) discipline is adopted while handling each queue.

Traffic model in this work includes mainly the following two procedures.

\begin{itemize}
  \item Packet Processing.  At each time step, node $k$ can process $C_k$ packets at most. For each of these packets, if node $k$ is not its destination, it is then delivered to the next stop toward its destination based on a specific routing algorithm. Otherwise, it is removed from the network.
  \item Packet Generation. At each time step, the network creates $R$ new packets with randomly chosen sources and destinations. For each packet, once its source and destination are determined, a path from the source to the destination is chosen based on a specific routing algorithm. If there are multiple paths, we choose one randomly. Note that the chosen path may be different at different time step, even with the same source-destination pair. The packet is then put at the end of the queue at its source node.
\end{itemize}

In order to characterize traffic congestion, we use the order parameter introduced in \cite{arenas2001communication}
\begin{equation}
\eta(R)=\lim_{t\rightarrow\infty}\frac{\langle\Delta L\rangle}{R\Delta t},
\end{equation}
where $L(t)$ is the total number of packets in the network at time $t$, $\Delta L=L(t+\Delta t)-L(t)$, $\langle \cdots \rangle$ indicates the average over time windows of width $\Delta t$. Actually, the order parameter indicates the traffic status in the macroscopic level. The traffic tie-up will be observed when the packet generation rate $R$ is sufficiently high. A critical value $R_{c}$ is thus expected to characterize the phase transition from free-flow to jamming. When $R<R_{c}$, the number of packets in the network is a constant, making $\eta$ be zero. While $R>R_{c}$, $\langle \Delta L \rangle$ , however, grows linearly with $\Delta t$. So $\eta$ is a constant larger than zero. In short, $R_{c}$ can be a measure of the traffic capacity.

In this paper, we analyze two typical routing criteria. One is the shortest path algorithm, which has been widely used in many transport networks. Given a single general network, Ling et al. showed analytically that the shortest path criterion can achieve the largest traffic capacity if the processing capacity of nodes is proportional to the corresponding algorithmic betweenness (i.e., the NUP allocation strategy is applied) \cite{xiang2013traffic}. The other is the efficient routing algorithm \cite{yan2006efficient}, which can avoid collapse of hub nodes by redistributing the traffic load from central nodes to other noncentral nodes. Such routing criterion can thus improve the traffic capacity greatly on scale-free networks if all nodes share the same processing capacity (i.e., the UNI allocation strategy is applied). For simplicity, we suppose that routing protocols on two originally isolated complex networks are identical and remain unchanged on newly formed interconnected complex networks.


\begin{figure*}[!hbtp]
\centering
\includegraphics[height=0.5\columnwidth,width=\columnwidth]{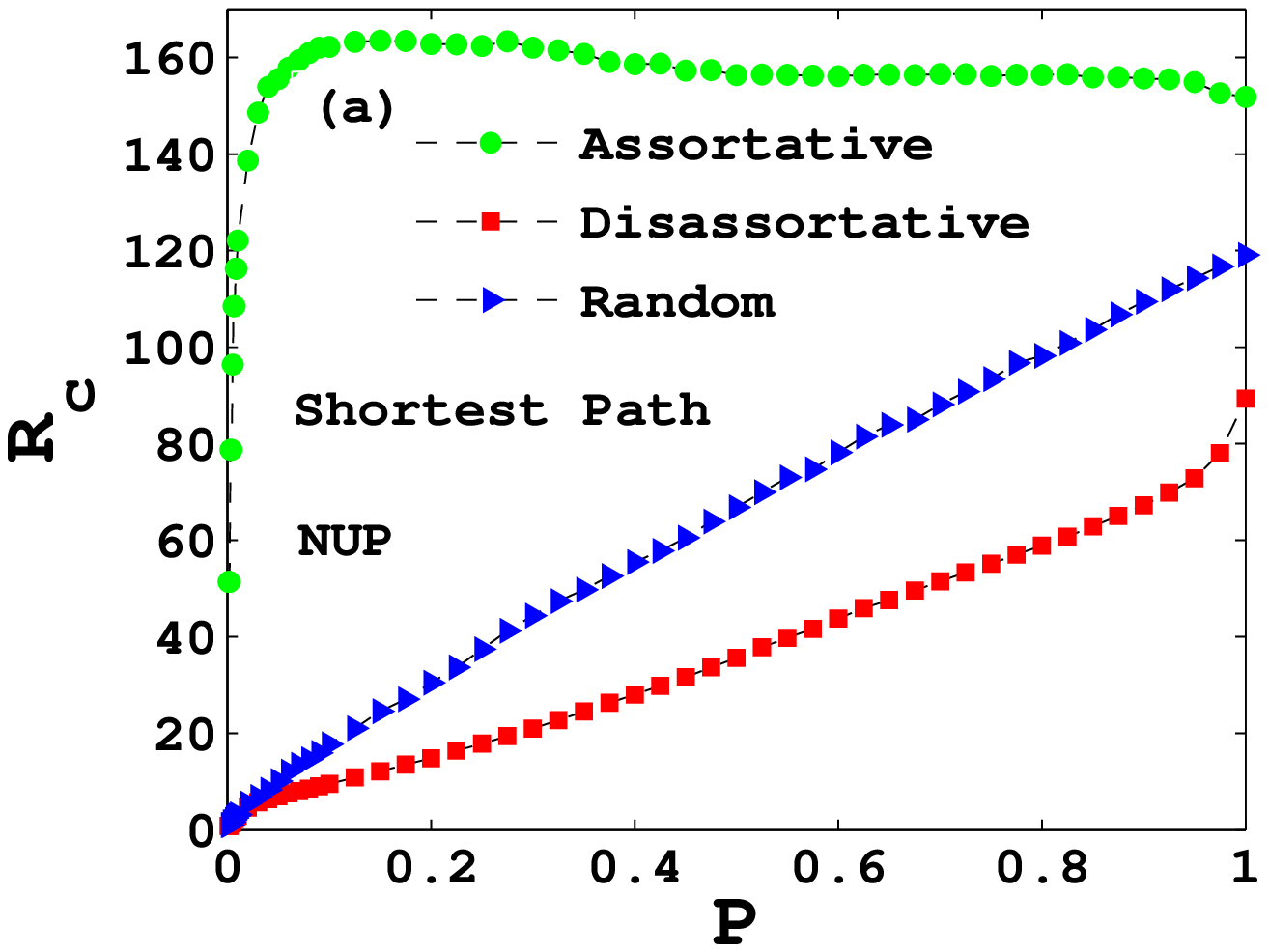}
\includegraphics[height=0.5\columnwidth,width=\columnwidth]{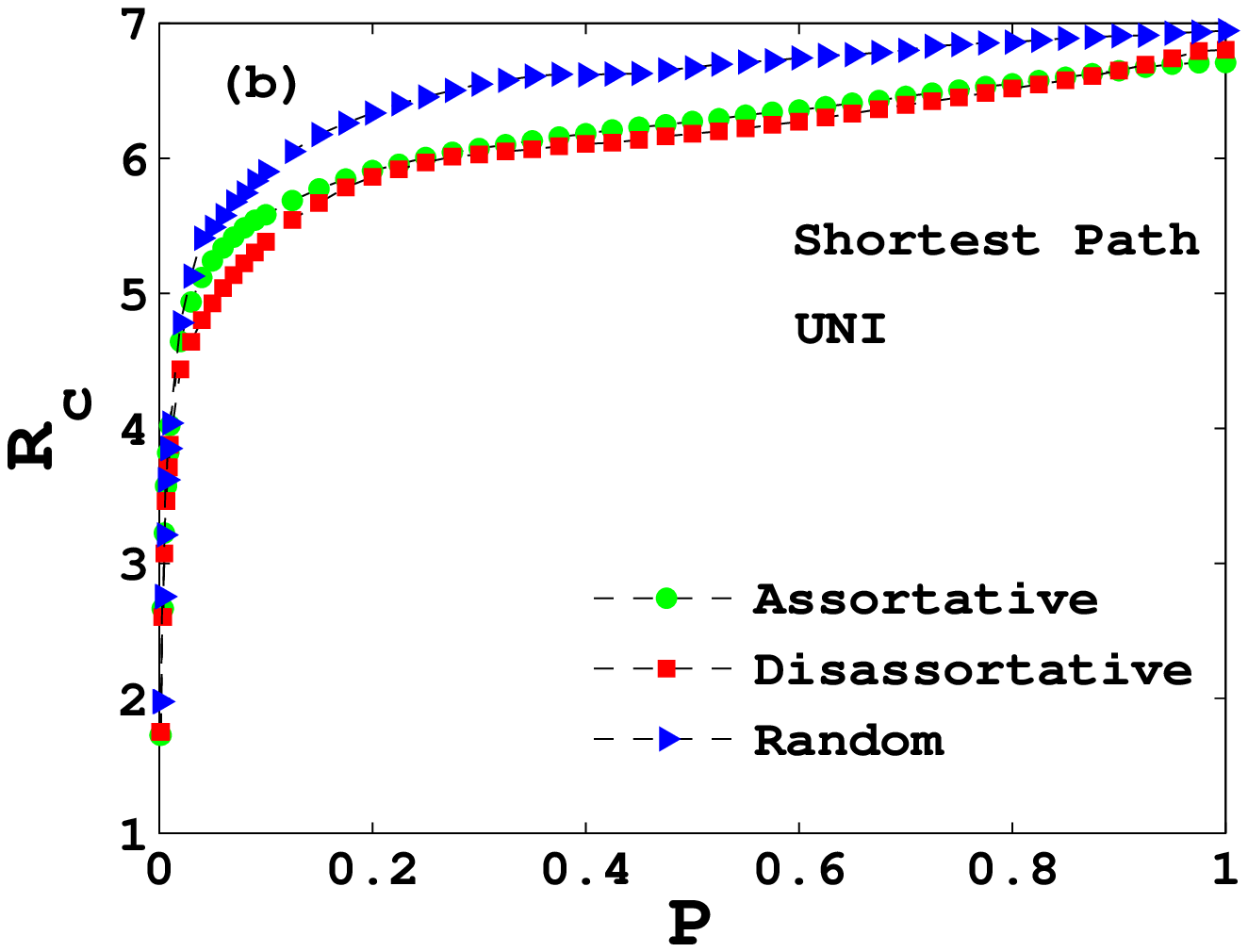}
\includegraphics[height=0.5\columnwidth,width=\columnwidth]{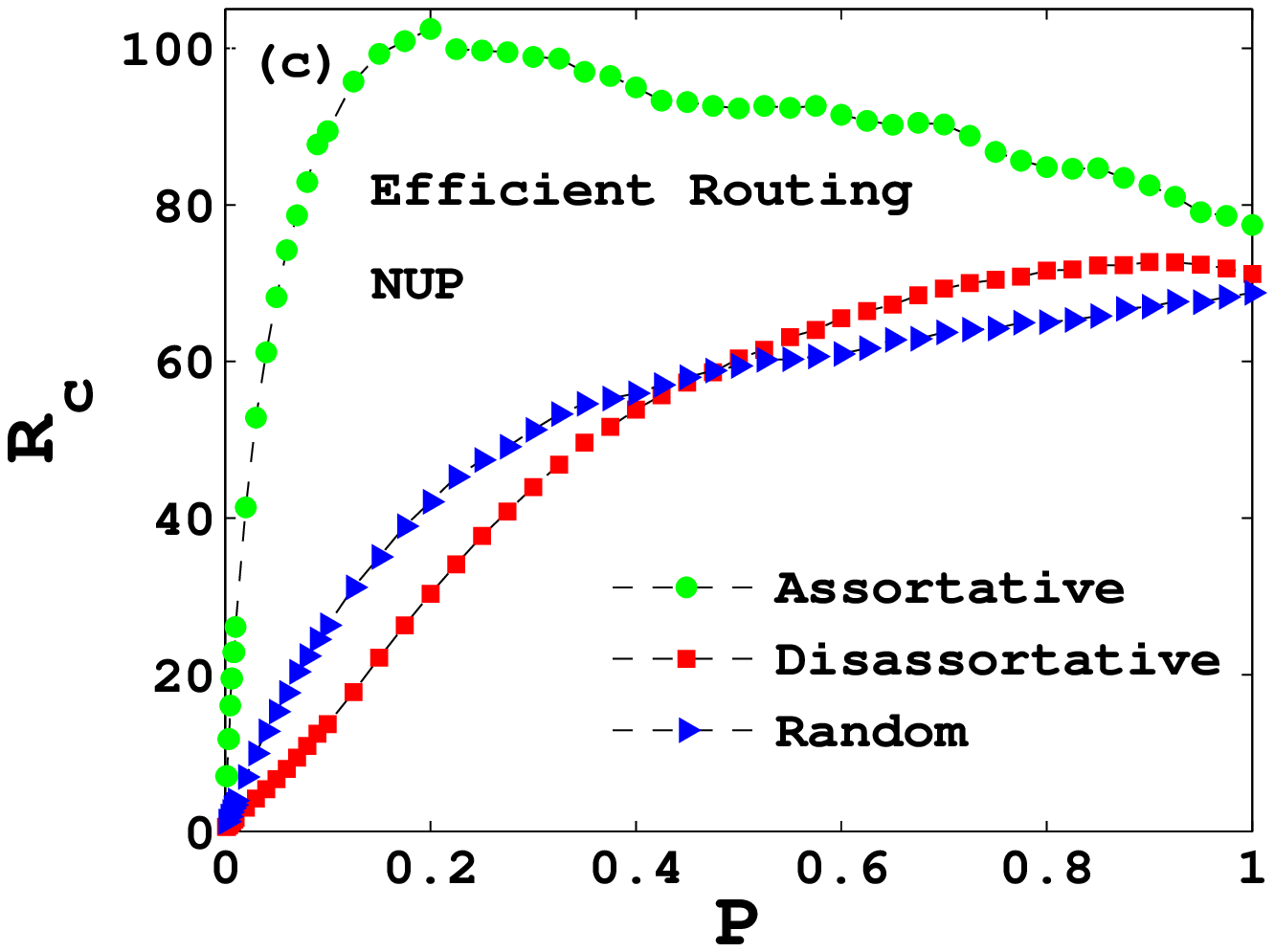}
\includegraphics[height=0.5\columnwidth,width=\columnwidth]{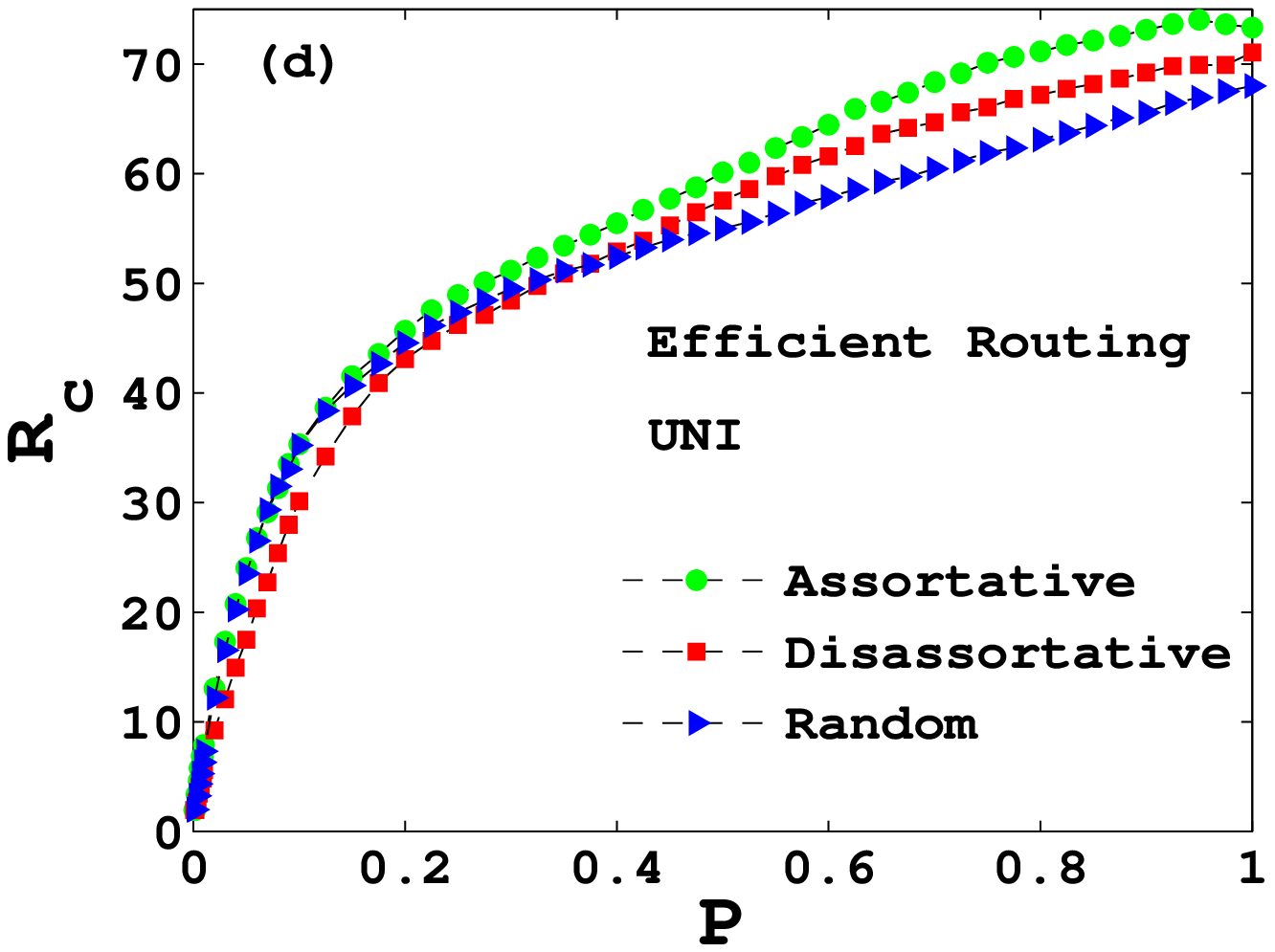}
\caption{(Color online) The traffic capacity $R_c$ as a function of the coupling probability $P$ for interconnected BA scale-free networks. The processing capacity is allocated based on node usage probability for shortest path (a) and efficient routing (c) criteria, and uniformly for shortest path (b) and efficient routing (d) criteria. $N_A=N_B=600$, and $\langle k_A \rangle=\langle k_B\rangle =4$. Each point is averaged over $50$ independent runs.}
\label{BA_BA}
\end{figure*}

\section{\label{sec:level3}Results}

In \cite{zhao2005onset}, Zhao et al. have given a theoretical estimation of the traffic capacity $R_c$ in isolated complex networks, which can be denoted as
\begin{equation}
\label{R_C}
R_c=\frac{N(N-1)}{(\frac{B_k}{C_k})_{max}},
\end{equation}
where $B_k$ and $C_k$ are the algorithmic betweenness and processing capacity of node $k$, respectively. This formula can apply to interconnected networks because interconnected links provide routes for traffic between two networks.

When the NUP allocation strategy is applied, we can revise Eq. \eqref{R_C} to estimate $R_c$ in interconnected networks with size $N_A+N_B$ by
\begin{equation}
\label{R_C_NUP}
\begin{split}
R_c=\frac{(N_A+N_B)(N_A+N_B-1)}{(\frac{B^{'}_i}{U_A(i)N_A},\frac{B^{'}_j}{U_B(j)N_B})_{max}},
\end{split}
\end{equation}
where $i$ $\in \{1,2, ..., N_A\}$, $j$ $\in \{1,2, ..., N_B\}$. $B^{'}_i$ and $B^{'}_j$ are the corresponding algorithmic betweenness of node $i$ in network $A$ and of node $j$ in network $B$ after two networks are interconnected.

If the UNI allocation scenario is adopted, the processing capacity of each node in interconnected networks is one.
We can also obtain
\begin{equation}
\label{R_C_UNI}
R_c=\frac{(N_A+N_B)(N_A+N_B-1)}{(B^{'}_k)_{max}},
\end{equation}
where $B^{'}_k$ is the algorithmic betweenness of node $k$ in newly formed interconnected networks.

\subsection{\label{sec:level2}Interconnected BA scale-free networks}

In this subsection, we will investigate how the interplay of interconnection and traffic resource allocation scenario affects traffic congestion for shortest path and efficient routing protocols in interconnected BA scale-free networks, respectively. In order to maintain tractability and facilitate analogy with real-world interconnected Internet AS-level graphs of South Korea and Japan to be analyzed in the next subsection, we carry out theoretical estimations on two BA scale-free networks of the equal size $N_A=N_B=600$ and average degree $\langle k_A \rangle=\langle k_B \rangle=4$.

Given two networks, fig. \ref{BA_BA} exhibits how the traffic capacity evolves with the coupling preference and probability under different routing criteria and traffic resource allocation scenarios.

Using Eq. \eqref{R_C_NUP}, subfigures \ref{BA_BA}(a) and (c) are obtained  based on shortest path and efficient routing strategies, respectively. Obviously, according to either one of the two routing protocols, assortative coupling outperforms greatly both diassortative and random coupling when the coupling probability $P$ increases from 0.001 to 1. What's more, the traffic capacity $R_c$ increases continuously with the coupling probability $P$ for disassortative and random coupling. The evolution of $R_c$ for assortative coupling with $P$, however, is away from such increasing trend. In particular, $R_c$ increases sharply at first and then decreases slightly with $P$. That is to say, there is an optimal coupling probability where the traffic capacity $R_c$ can achieve the maximum. These phenomena can be explained as follows.

First, different coupling preferences have different effects on the congestion status. According to Eq. \eqref{R_C_NUP}, in order to alleviate traffic congestion, the adjustment of interconnected links is supposed to balance the traffic load of all nodes based on the node processing capacity distribution. Furthermore, two ends of each interconnected link share the same load caused by traffic across two networks. Thus, if two ends of an interconnected link have similar processing capacity, traffic congestion can be mitigated to the largest extent. In our model, two BA scale-free networks share the same size and average degree, assortative coupling can thus meet such requirement better than the other two coupling patterns.

Second, the interplay between coupling preference and coupling probability makes the shape of curves in subfigures \ref{BA_BA}(a) and (c) different. Regarding assortative coupling, when few links are attached between two networks, the newly generated traffic load between two networks is mainly accumulated on nodes with large processing capacity. The accumulation of load makes such nodes congested more easily. In this sense, more links help to distribute such load and thus alleviate traffic congestion. But with the continuous increment of interconnected links, those nodes with small processing capacity have to accommodate more load triggered by traffic between two networks. When the ratio between the total load (caused by traffic both within and across two networks) and the processing capacity of nodes with small processing capacity exceeds that of nodes with large processing capacity, the former will trigger traffic congestion at first. Thus, more links mean that more traffic load is distributed to nodes with small processing capacity in general. That is to say, more severe traffic congestion will occur accordingly. However, for both disassortative and random coupling, due to the heterogeneity of scale-free networks and coupling mechanism, two nodes both with large processing capacity in respective networks can hardly be interconnected. Consequently, traffic congestion is always caused by nodes with small processing capacity in all the range of the coupling probability. Therefore, more links facilitate the even distribution of traffic load.

Using Eq. \eqref{R_C_UNI}, we can also have subfigures \ref{BA_BA}(b) and (d) based on the UNI resource allocation strategy. As one can see, different from the NUP allocation strategy, here three different types of coupling preferences achieve almost the same traffic capacity with the increase of coupling probability. Meanwhile, for each type of coupling preference, the traffic capacity increases continuously. In other words, although the disparity of locations of interconnected links can undoubtedly trigger traffic load distribution adjustments of many nodes, the largest traffic load of all nodes in interconnected networks is immune to such variation to some extent. As regards with the coupling probability, more interconnected links mean more traffic paths between two originally isolated networks. This makes traffic load distribution more even and thus reduces the largest traffic load. In accordance with Eq. \eqref{R_C_UNI}, traffic performance will be improved continuously. Therefore, if the processing capacity of all nodes is identical, we can choose the coupling preference at random from the perspective of controlling traffic congestion in interconnected networks.

Putting four subfigures together, we can have some further insights. First, with respect to traffic performance for assortative coupling, subfigures (a) and (b) are respectively the best and worst among four subfigures. This is in agreement with previous findings in isolated complex networks \cite{xiang2013traffic}. Second, for both shortest path and efficient routing criteria, the NUP strategy can achieve better traffic performance compared to the UNI strategy. Thirdly, the shortest path protocol outperforms the efficient routing protocol when the NUP strategy is adopted as exhibited in subfigures (a) and (c). Whereas if the UNI strategy is used, the efficient routing protocol is better than the shortest path protocol as shown in subfigures (b) and (d). These results teach us that we have to clarify the resource allocation strategy before we compare two routing protocols.

It's worthy to mention that the aforementioned results are obtained using the formulas (\ref{R_C_NUP}) and (\ref{R_C_UNI}). In addition, we have checked the corresponding simulation results on interconnected BA scale-free networks. They are also in good line with the theoretical estimation results.

\subsection{\label{sec:level2}Interconnected Internet AS-level graphs of South Korea and Japan}

\begin{figure}[!hbtp]
\centering
\includegraphics[height=\columnwidth,width=\columnwidth]{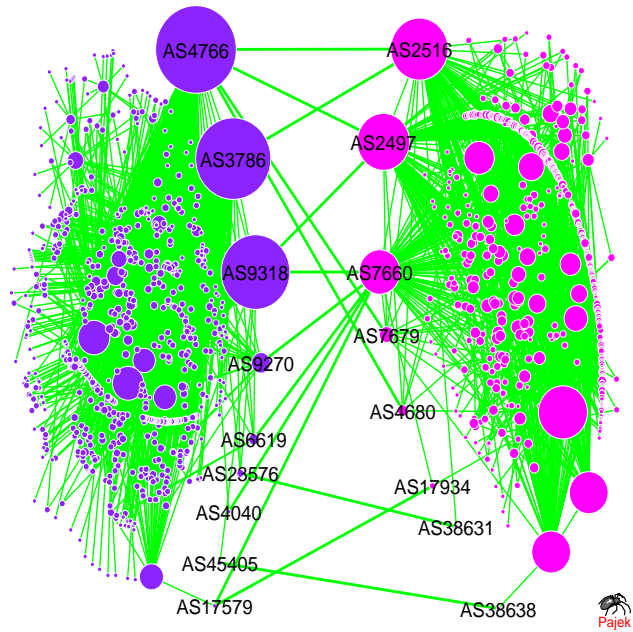}
\caption{(Color online) Visualization of interconnected Internet AS-level topologies of South Korea (left) and Japan (right) labeled as SK and JP, respectively. The size of nodes represents their internal degree, and label of nodes is autonomous system number \cite{ASNAME}. Interconnected links across SK and JP are listed and labeled in Table \ref{tab:table1}. $N_{SK} = 677$, $N_{JP} = 509$, $\langle k_{SK} \rangle \approx 3.65$ and $\langle k_{JP} \rangle \approx 4.40$. Visualization: Pajek (Batagelj and Mrvar, 2013).}
\label{VISUALIZATION}
\end{figure}

\begin{figure}[!hbtp]
\centering
\includegraphics[height=\columnwidth,width=\columnwidth]{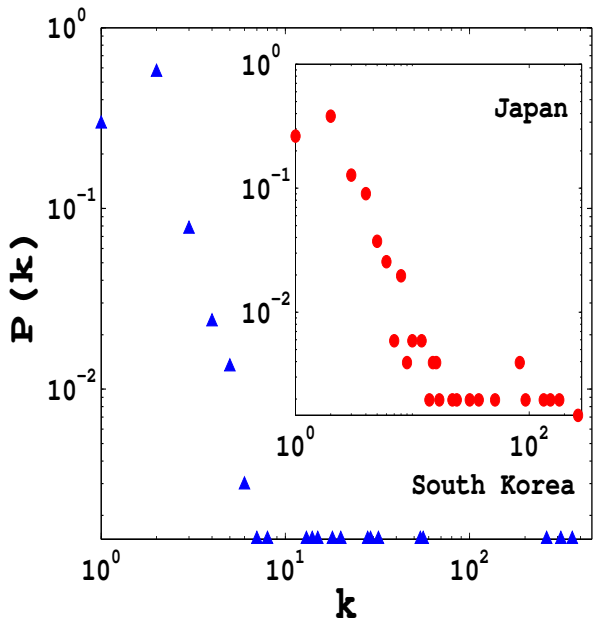}
\caption{(Color online)  The distribution function of internal edges for networks SK and JP. (main figure) Internet AS-level topologies of South Korea, with $N_{SK}=677$ nodes and average internal connectivity $\langle k_{SK} \rangle \approx 3.65$. (inset) Internet AS-level topologies of Japan, with $N_{JP}=509$ and $\langle k_{JP} \rangle \approx 4.40$.}
\label{DEGREE}
\end{figure}

\begin{table}[!t]
\renewcommand{\arraystretch}{1.3}
\caption{\label{tab:table1}%
Fourteen interconnected links are labeled from $1$ to $14$, and the corresponding two ends of each link are listed. More information can be available in \cite{ASNAME}. }
\centering
\begin{tabular}{|c|c|c||c|c|c|}
\hline
 Label & South Korea & Japan  & Label & South Korea & Japan  \\
\hline
1 & AS3786 & AS2516  & 8 & AS9270 & AS7660   \\
\hline
 2 & AS4040 & AS7660  & 9 & AS9318 & AS2497   \\
\hline
 3 & AS4766 & AS2497  & 10 & AS9318 & AS7660  \\
\hline
 4 & AS4766 & AS2516  & 11 & AS17579 & AS7660  \\
\hline
5 & AS4766 & AS4680  & 12 & AS17579 & AS17934  \\
\hline
 6 & AS4766 & AS7679  & 13 & AS23576 & AS38631  \\
\hline
7 & AS6619 & AS7660  & 14 & AS45405 & AS38638  \\
\hline
\end{tabular}
\end{table}

\begin{figure*}[!hbtp]
\centering
\includegraphics[height=0.5\columnwidth,width=\columnwidth]{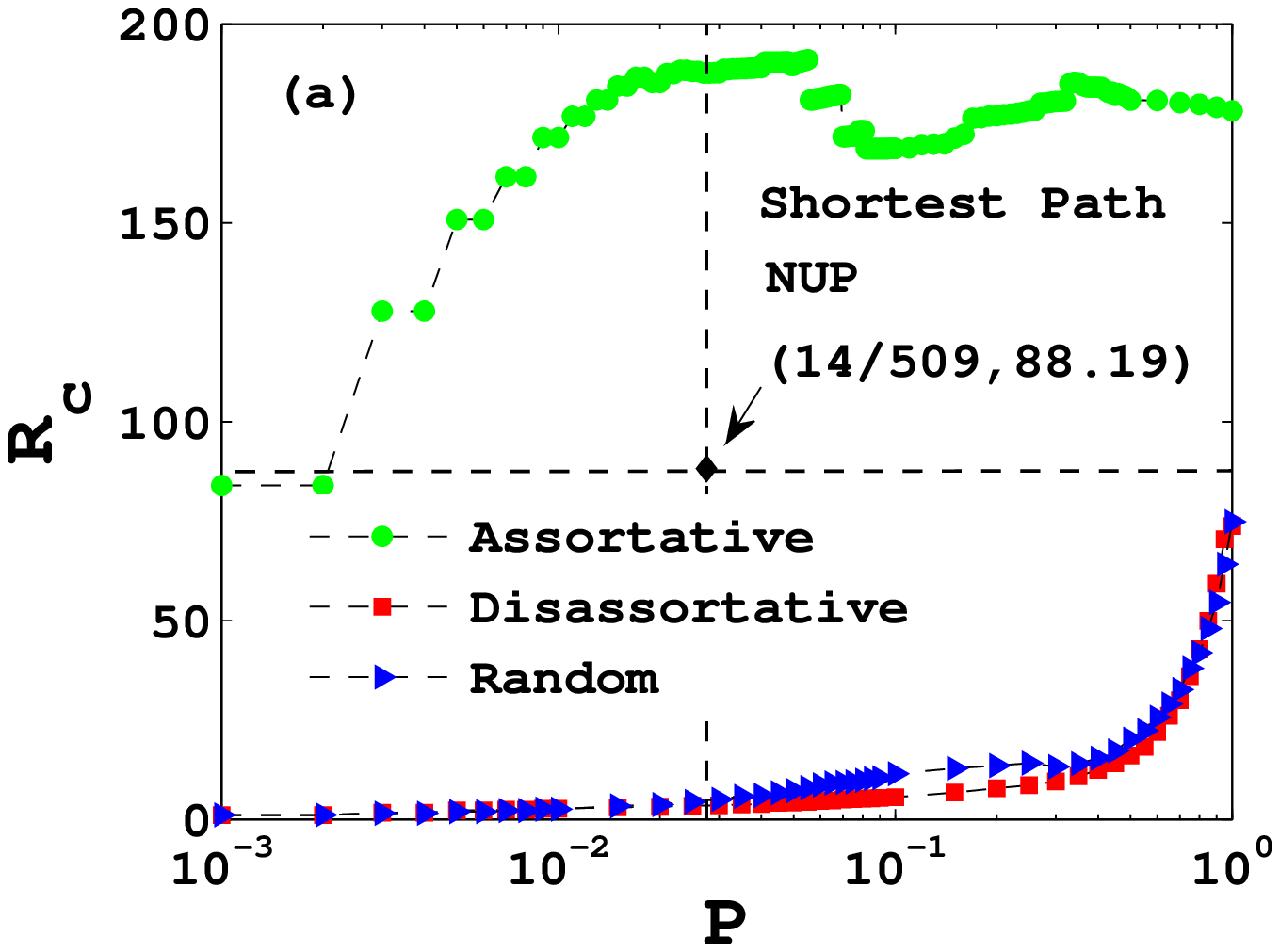}
\includegraphics[height=0.5\columnwidth,width=\columnwidth]{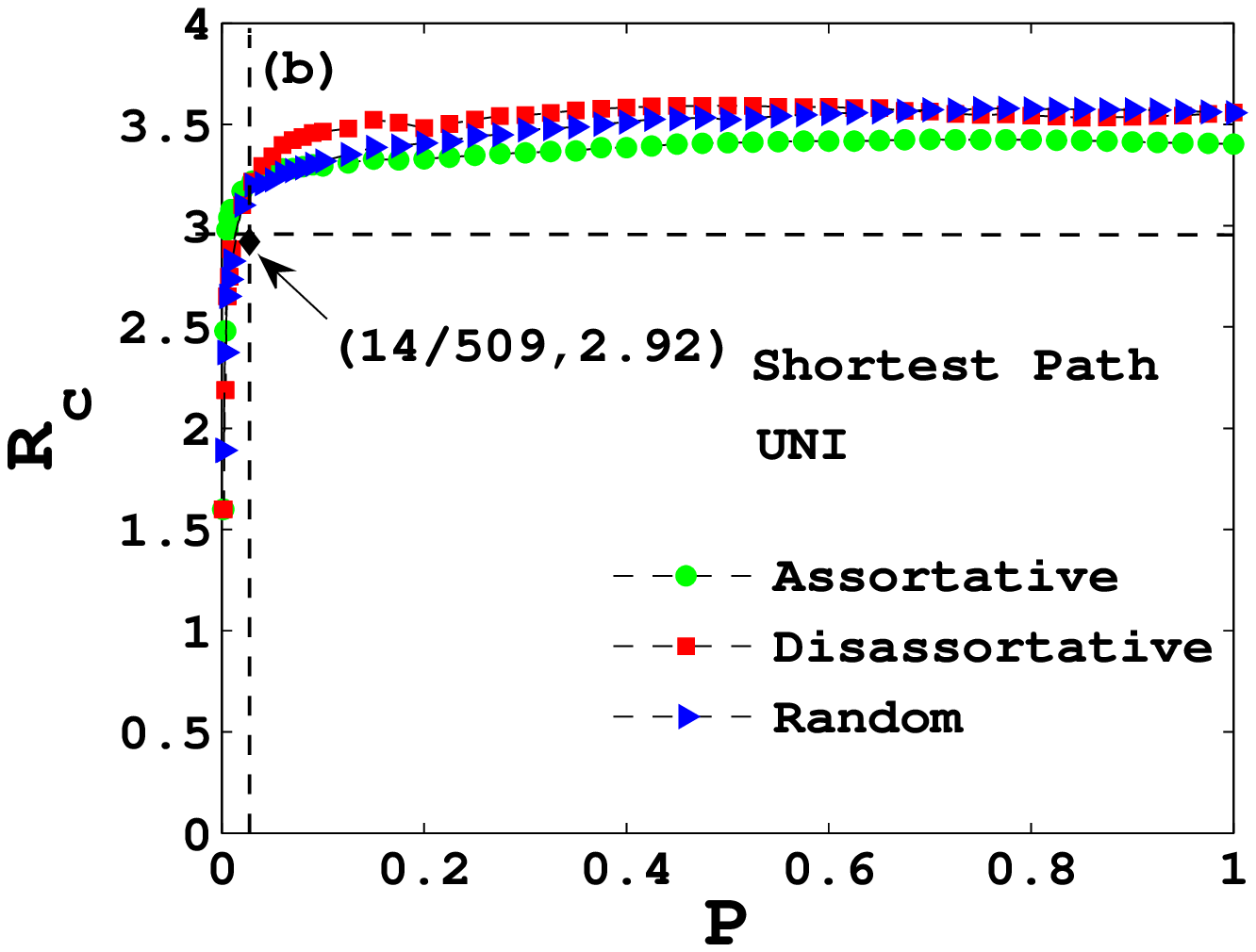}
\includegraphics[height=0.5\columnwidth,width=\columnwidth]{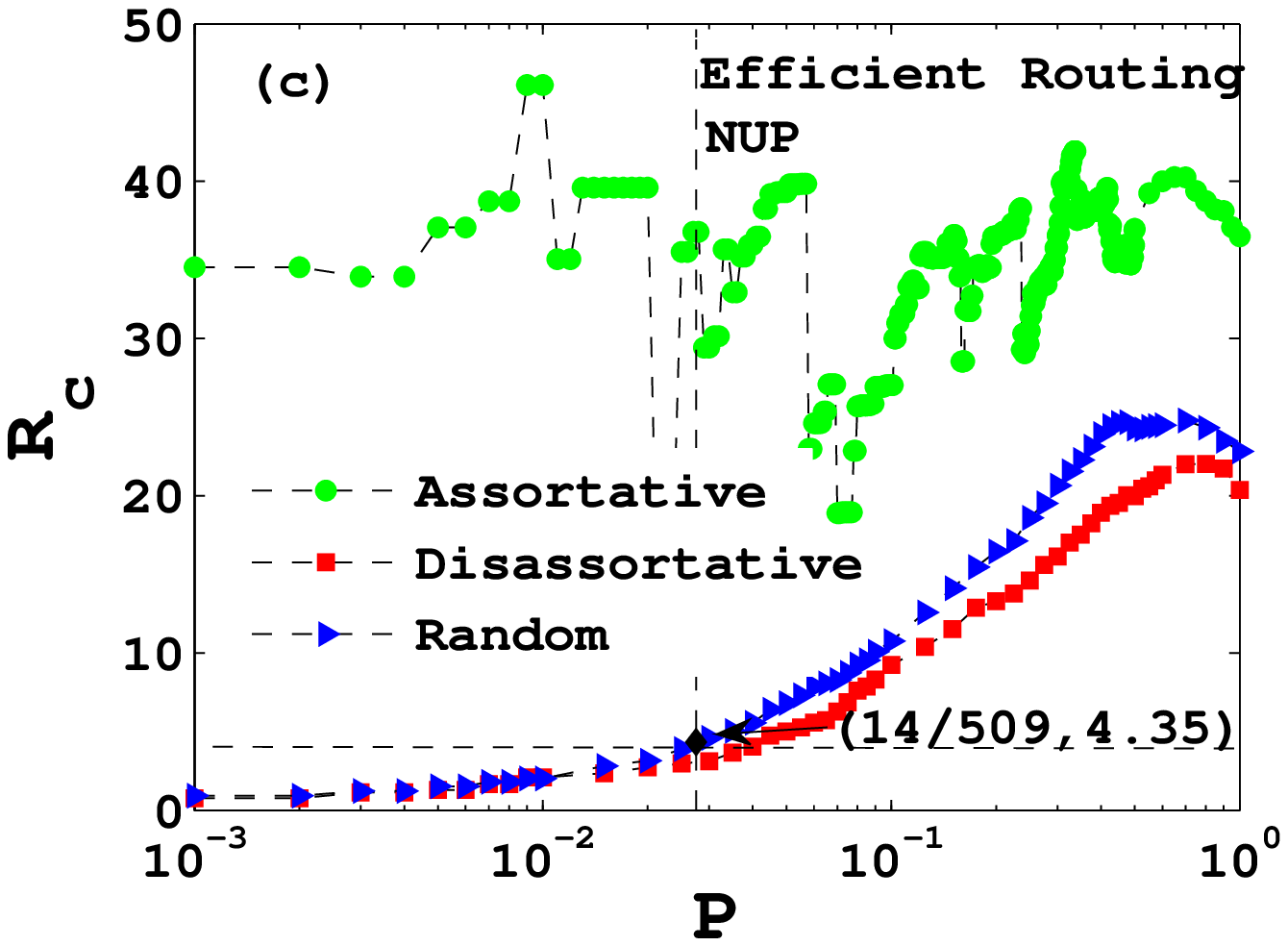}
\includegraphics[height=0.5\columnwidth,width=\columnwidth]{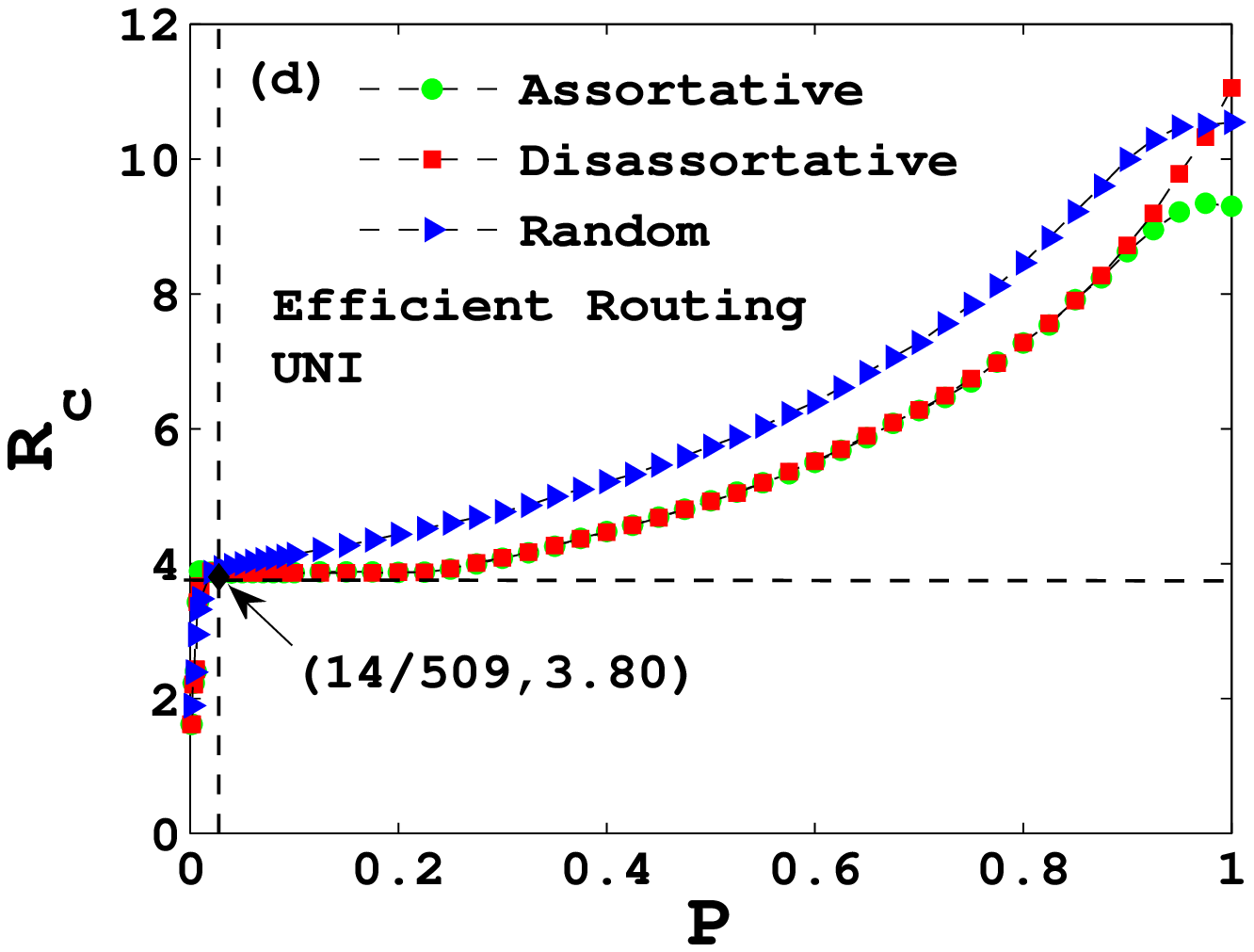}
\caption{(Color online) Evolution of the traffic capacity $R_c$ versus the coupling probability $P$ for interconnected Internet AS-level topologies of South Korea and Japan. The processing capacity is allocated based on node usage probability for shortest path (a) and efficient routing (c) criteria, and uniformly for shortest path (b) and efficient routing (d) criteria. The corresponding intersection of two dashed lines represents the real situation. $N_{SK}=677$, $N_{JP} = 509$, $\langle k_{SK} \rangle \approx 3.65$ and $\langle k_{JP} \rangle \approx 4.40$. Each point is averaged over $50$ independent runs.}
\label{JP_KR}
\end{figure*}

Many real-world networks evolve by interconnecting originally isolated subsystems. As such, traffic congestion is expected to be influenced greatly by these interconnected links. As an example, we here focus on two interconnected components of the Internet at autonomous system level in South Korea and Japan, which we label as SK and JP respectively (fig. \ref{VISUALIZATION}). These two networks are also connected to networks of other countries or regions, which we ignore in this paper. We obtain topological data from the Autonomous System Ranking provided by the Cooperative Association for Internet Data Analysis (dataset version: 2013-04-01) \cite{CAIDA}. Networks SK and JP are of sizes $N_{SK}=677$ and $N_{JP}=509$. They have rather different average internal degrees ($\langle k_{SK} \rangle \approx 3.65$ and $\langle k_{JP} \rangle \approx 4.40$, respectively) but both exhibit a power-law distribution of internal links as shown in fig. \ref{DEGREE}. It is also found that these two networks are sparsely interconnected by just fourteen external edges which can be found in detail in Table \ref{tab:table1}.

The congestion thresholds for the interconnected Internet graphs are shown as the corresponding intersections of dashed lines in four subfigures in fig. \ref{JP_KR}, with different routing protocols and resource allocation strategies, respectively. It is found that the shortest path protocol along with the NUP allocation strategy (subfigure \ref{JP_KR}(a)) achieves the highest $R_c$, whereas the shortest path protocol along with the UNI allocation strategy (subfigure \ref{JP_KR}(b)) performs the worst. This result is in perfect agreement with our finding with idealized interconnected BA networks model.

To verify our previous findings about the effect of coupling preference/probability on traffic congestion in the idealized network model, we conduct the same estimations on networks SK and JP. Different from the idealized model of two networks with the same size and average degree, here the network sizes $N_{SK} \neq N_{JP}$. So we modify the definition of coupling probability $P$ as
\begin{equation}
P=\frac{N_{il}}{(N_{SK},N_{JP})_{min}}.
\end{equation}
Since one node has one interconnected link at most, $P$ is still in the interval $[0,1]$. In order to show the effect of coupling preference, it is assumed that fourteen links are not established yet. In other words, we only keep all the links within each individual network, and add interconnected links with the specific coupling preference. Specifically, in the disassortative coupling case, it is assumed that nodes in networks SK and JP are sorted in the descending and ascending order of load, respectively.

\begin{figure}[!hbtp]
\centering
\includegraphics[height=0.5\columnwidth,width=\columnwidth]{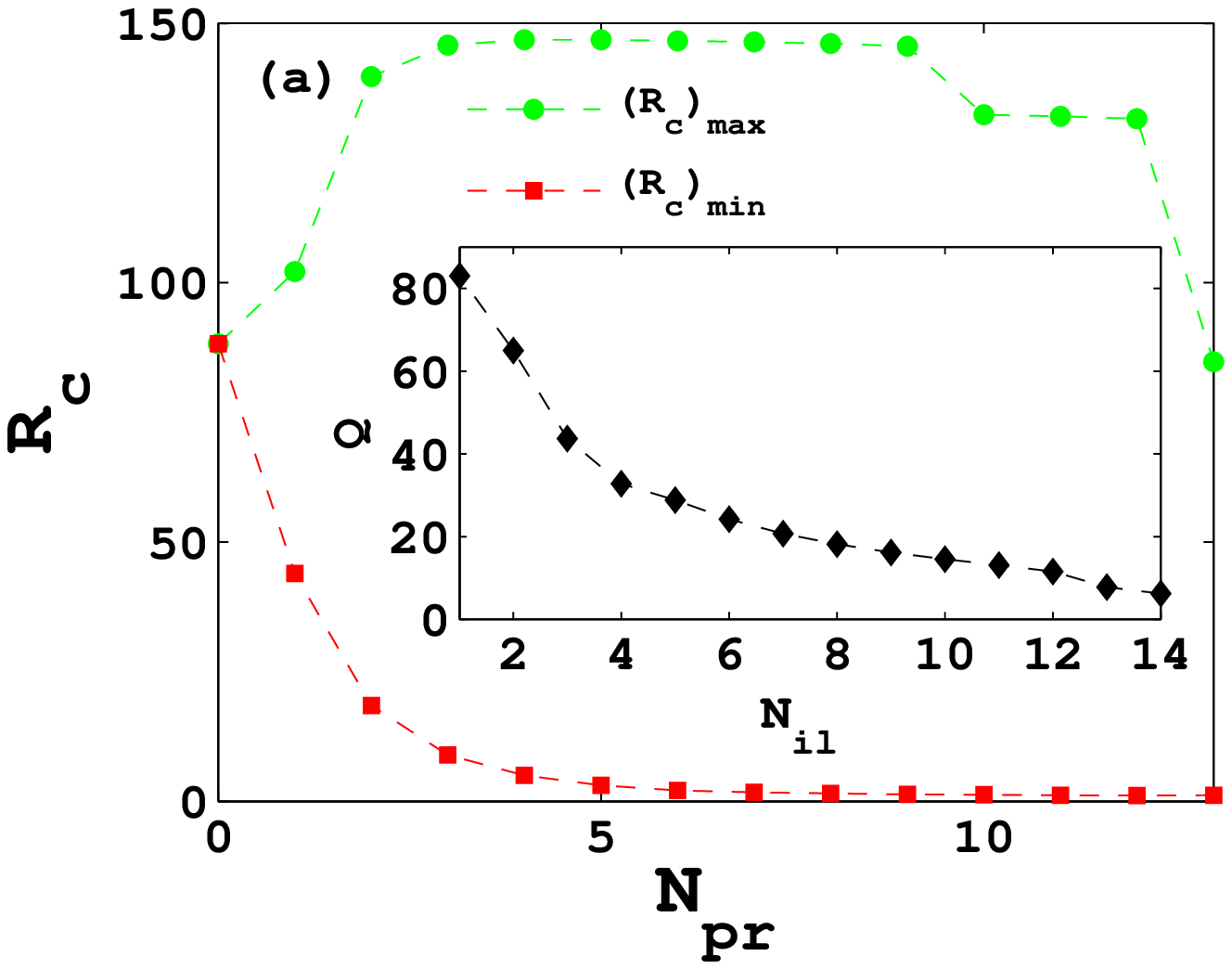}
\includegraphics[height=0.5\columnwidth,width=\columnwidth]{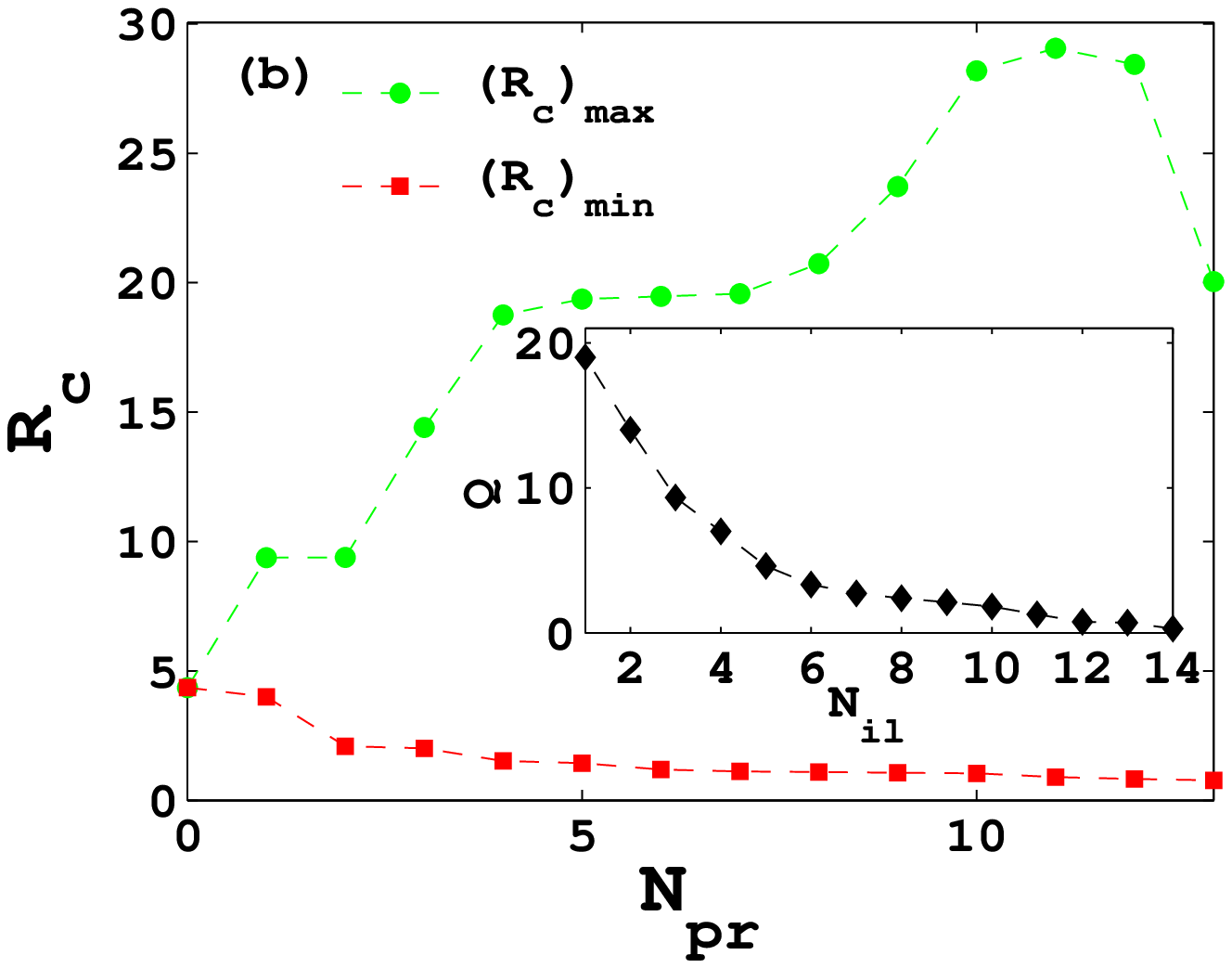}
\caption{(Color online) The maximum and minimum of the traffic capacity $R_c$ versus the number of pruned links $N_{pr}$ between networks SK and JP for shortest path (a) and efficient routing (b) protocols. The corresponding inset shows the evolution of Q with the number of interconnected links $N_{il}$. }
\label{RC_NR}
\end{figure}

Subfigures \ref{JP_KR}(a) and (c) demonstrate that assortative coupling can outplay both disassortative and random coupling. In subfigure \ref{JP_KR}(a), say, the minimum of the traffic capacity for assortative coupling is even larger than the maximum of the traffic capacity for the other two coupling patterns. Besides, for disassortative and random coupling, $R_c$ increases with $P$. There is also a critical coupling probability for assortative coupling in general. In subfigure \ref{JP_KR}(c), however, critical coupling probability can be found in both disassortative and random coupling. This is different from what we have obtained in the idealized model. Furthermore, the trend of $R_c$ for assortative coupling is surprisingly irregular. These disparities to some extent arise from the interplay of difference of size and average degree between networks SK and JP and few super high-degree nodes. Nodes $AS4766$ and $AS3786$ in network SK, for example, have respectively $361$ and $313$ internal links which are more than half the network size. Although the efficient routing protocol is adopted here to bypass hub nodes, a large amount of traffic load is still accumulated on these super high-degree nodes. That is to say, they are inevitable for traffic between most of other noncentral nodes. Despite such differences between real-world network and the idealized network model, assortative coupling is optimal if the processing capacity of nodes is allocated based on node usage probability.

Similarly to what we have concluded from subfigures \ref{BA_BA}(b) and (d), the traffic capacity remains unaffected by the coupling preference as shown in subfigures \ref{JP_KR}(b) and (d). Particularly, $R_c$ increases sharply at first and then remains steady or increases slowly. In either case, the coupling preference makes no difference in terms of alleviating traffic congestion. This is in agreement with analysis in the aforementioned idealized model. The intersection of two dashed lines in these two subfigures show the actual traffic capacity of interconnected AS-level graphs of South Korea and Japan. As one can see, although these fourteen links don't follow any one of three types of coupling preferences here, the corresponding traffic capacity are surprisingly similar. This result thus further confirms our conclusion.

As we can see from subfigures \ref{JP_KR}(a) and (c), fourteen established links between networks SK and JP are not optimal. Thus, they can be modified to mitigate traffic congestion and improve the traffic capacity in interconnected networks. Actually, previous studies have showed that both link addition and pruning can accomplish such goals \cite{zhang2007enhancing,huang2010effective}.
Considering the economical and technical cost, link pruning is a better strategy than link addition. Hence, we gradually remove these links one by one and find the respective maximum and minimum of the traffic capacity for each step, namely $(R_c)_{max}$ and $(R_c)_{min}$. This procedure is shown in fig. \ref{RC_NR}. It demonstrates the evolution of the traffic capacity with the number of pruned links $N_{pr}$ for shortest path and efficient routing protocols, respectively. As one can see, the traffic capacity is sensitive to both the position and number of pruned links. Particularly, when more interconnected links are pruned, $(R_c)_{max}$ increases at first and then decreases slightly for both shortest path and efficient routing algorithms. Nonetheless, $(R_c)_{min}$ decreases continuously with $N_{pr}$. For the shortest path protocol, when $N_{pr}=9$, we can obtain the largest traffic capacity through just five remaining links labeled as $\{1,3,4,9,10\}$. For the efficient routing criterion, when $N_{pr}=12$, the largest traffic capacity can be available through just two remaining  links labeled as $\{1,9\}$.

In real-world practice, however, the laying and maintenance of cables between different countries or regions are usually costly. Every cable is thus expected to contribute to the improvement of traffic performance to the largest extent. In this sense, we introduce a quantity $Q$ to reflect such concern, which can be defined as $Q=\frac{R_c}{N_{il}}$. We here focus on the evolution of $(R_c)_{max}$, thus $Q=\frac{(R_c)_{max}}{N_{il}}$.
The corresponding insets exhibit the relation between $Q$ and the number of interconnected links $N_{il}$. $Q$ decreases monotonously with $N_{il}$ for two kinds of routing strategies. Interestingly, this phenomenon is in agreement with the law of diminishing marginal utility in economics \cite{hirschey2008fundamentals}. From the perspective of maximizing $Q$, $N_{il}=1$ is optimal. We find that links labeled as $\{4\}$ and $\{9\}$ are the remaining ones for shortest path and efficient routing protocols, respectively. Furthermore, no matter maximizing $R_c$ or $Q$, interconnected link labeled as $\{4\}$ and $\{9\}$ cannot be pruned for respective protocols. Two ends of unpruned link $\{4\}$ are hub nodes. This complies with the shortest path protocol. However, for the efficient routing protocol, two hub nodes, namely AS9318 in network SK and AS2497 in network JP, are unavoidable even though such protocol is designed to bypass hub nodes. This is what real-world networks differentiate from the idealized model. Therefore, based on this work, our future work is to explore possible effective routing protocols in real-wold interconnected networks.

Similarly to subsection A, we have conducted the simulation on interconnected Internet AS-level graphs of South Korea and Japan, which supports the theoretical estimations very well.


\section{\label{sec:level4}Conclusions}

We have investigated traffic congestion in interconnected complex networks in this paper. For interconnected BA scale-free networks, it is found that assortative coupling is more helpful to mitigate traffic congestion than both disassortative and random coupling if the NUP scenario is applied. Particularly, the optimal coupling probability can be found for assortative coupling.
Whereas if all nodes share the same processing capacity, traffic congestion isn't swayed by the coupling preference. In this case, traffic congestion can be alleviated while more interconnected links are attached. Similar results apply to interconnected Internet AS-level graphs of South Korea and Japan. According to these results, we give some practical proposals for optimization of interconnected links in interconnected AS-level graphs. Altogether, this paper provides a reasonable approach to layout of interconnected links when traffic congestion is taken into account. Moreover, our work will attract researchers to design more efficient individualized transport protocols for real-world interconnected networks.

\begin{acknowledgments}
This work was supported by Hong Kong Research Grants Council under Grant No. PolyU 5262/11E and the National Natural Science Foundation of China under Grant No. 61174153. Jiajing Wu was supported by the Hong Kong PhD Fellowship Scheme.
\end{acknowledgments}

\bibliographystyle{apsrev}
\bibliography{Traffic_Networks}

\end{document}